\documentclass[preprint,showpacs,preprintnumbers,amsmath,mprb]{revtex4-1}

\usepackage{graphicx}                % Include figure files
\usepackage{dcolumn}                 % Align table columns on decimal point
\usepackage{bm}                      % bold math
\usepackage{hyperref}
\usepackage{color}
\usepackage{amsfonts}
\usepackage{upgreek}

\begin{document}

\title{Post-selected indistinguishable single-photon emission from the Mollow triplet sidebands of a resonantly excited quantum dot}

\author{S. Weiler$^{1} \ast$, D. Stojanovic$^{1}$, S.~M.~Ulrich$^{1}$, M. Jetter$^{1}$,
 and P. Michler$^{1}$}

\affiliation{$^{1}$Institut f\"ur Halbleiteroptik und Funktionelle
Grenzfl\"achen, Universit\"at Stuttgart, Allmandring 3, 70569
Stuttgart, Germany,}

$^{\star}$~Corresponding author: s.weiler@ihfg.uni-stuttgart.de;
Homepage: http://www.ihfg.uni-stuttgart.de

%\date{\today}

\begin{abstract}

Applying high power continuous-wave resonant s-shell excitation to
a single self-assembled InGaAs quantum dot, we demonstrate the
generation of post-selected single-indistinguishable photons from
the Mollow triplet sidebands. A sophisticated spatial filtering
technique based on a double Michelson interferometer enabled us to
separate the spectrally close lying individual Mollow components
and perform almost background free two-photon interference
measurements. Showing high consistency with the results of an
independent determination of the emission coherence of $\sim 250
\pm 30 \, ps$, our analysis reveals a close to ideal visibility
contrast of up to $97 \%$. Due to their easy spectral tunability
and their distinct advantage of cascaded photon emission between
the individual Mollow sidebands, they resemble a versatile tool
for quantum information applications.

\end{abstract}



\maketitle

Investigations on the emission characteristics of single-photons
sources such as trapped atoms \cite{McKeever.Boca:2004}, molecules
\cite{Lounis.Moerner:2000} and quantum dots
\cite{Michler.Kiraz:2000} are a rich field of science, especially
in the focus of fundamental applications in quantum information
technologies \cite{Kimble:2008,OBrien.Furusawa:2009}. The
integratability of quantum dots (QDs) serving as deterministic
single-photon sources, with the additionally ability to fulfill
the often required precondition of emitting indistinguishable
photons, make these solid state based "artificial atoms"
attractive for intense research.

Several previous works on indistinguishable photon emission from
single QDs relying on Hong-Ou-Mandel type two-photon interference
at a beamsplitter \cite{Hong.Ou:1987} under non-resonant emitter
state preparation have revealed visibilities up to $~82 \%$
\cite{Santori.Fattal:2002,Weiler.Ulhaq:2010,Varoutsis.Laurent:2005}.

These preceding studies have paved the way for several profound
improvements for the application of QDs in quantum information
technology, e.g. the proof of indistinguishable photon emission
from spatially separate QDs
\cite{Flagg.Muller:2010,Patel.Bennett:2010}, as well as from
photons emitted from mutually different sources. Here the bosonic
nature in two-photon interference between QD emission and light
created by parametric down-conversion has been successfully
verified \cite{Polyakov.Muller:2011}. Another major achievement in
terms of implementation has been the proof of indistinguishable
photon emission from site-controlled single QDs
\cite{Jons.Atkinson:2013}.

Other ambitious developments focus on the creation of maximally
indistinguishable photons. By employing strictly resonant
emitter-state preparation, the visibility-limiting relaxation of
electron-hole pairs from higher energetic QD states into its
ground state is prohibited. Exploring the different regimes of
resonance fluorescence, highly indistinguishable photons have been
verified especially in the low excitation power limit so called
Heitler regime
\cite{Matthiesen.Vamivakas:2012,Nguyen.Sallen:2012,Matthiesen.Geller:2012}.
Here the absorption and emission process becomes one coherent
event and the scattered photons are free from any dephasing. In
the regime close to emission saturation, non-classical two-photon
interference has revealed indistinguishable photons both for
continuous wave (cw) excitation \cite{Ates.Ulrich:2009}, yielding
a post-selected visibility contrast of $V_{HOM} = 90 \%$, as well
as for triggered resonant excitation with $\pi$-pulses
\cite{He.He:2013} ($V_{HOM}=97\%$). The high-power dressed state
regime, experimentally accessed with QDs by Muller et al.
\cite{Muller.Flagg:2007} for the first time, reveals the Mollow
triplet as the characteristic hallmark of strongly driven
resonance fluorescence. Single and cascaded photon emission
between the Mollow sidebands has also been experimentally verified
recently \cite{Ulhaq.Weiler:2012}.

Inspired by these investigations, the aim of the current work is
the verification and analysis of indistinguishable photons from
the individual Mollow triplet sidebands with the distinct
advantage of controlled spectral tunability via laser excitation
power variation and/or resonance tuning.\\

In this letter we provide experimental evidence of post-selected
indistinguishable photon emission from individual Mollow triplet
sidebands of a cw resonantly excited single  InGaAs QD for
different pump powers, i.e. Rabi splittings. The coherence times
of the sidebands extracted from independent Mollow triplet power
series are, together with the relevant bunching and decay
timescales derived from $g^{(2)}(\tau)$-type second-order
auto-correlation measurements found to be in high consistency with the two-photon interference data.\\

The planar sample employed for the measurements in this work is
grown by metal-organic vapor phase epitaxy. The self-assembled
In(Ga)As QDs are embedded in a GaAs $\lambda$-cavity, sandwiched
between 29 (4) periods of $\lambda/4$-thick AlAs/GaAs layers as
bottom (top) distributed Bragg reflectors. For our experimental
investigations, the sample is kept in a Helium-flow cryostat
providing a highly stable temperature of $T = 5.0 \pm 0.5$~K.
Suppression of parasitic laser stray-light is achieved by use of
an orthogonal geometry between QD excitation and emission
detection in combination with polarization suppression and spatial
filtering via a pinhole in the detection path. The QDs are
resonantly excited by a narrow-band ($\approx 500$~kHz)
continuous-wave Ti:Sapphire ring laser. For high-resolution
spectroscopy (HRPL) of micro-photoluminescence ($\mu$-PL) we
employ a scanning Fabry-P\'{e}rot interferometer with $\Delta
E^{\rm HRPL}_{\rm res}< 1\,\mu{\rm eV}$ as described earlier
\cite{Ates.Ulrich:2009,Ulhaq.Weiler:2012,Ulrich.Ates:2011}.\\

% insert Fig. 1 here

A high degree of photon indistinguishability relies on the
spectral purity of the detected photons. Therefore we utilize a
double Michelson interferometer (motorized linear stages with a
stage resolution of $20\,nm$) in combination with a spectrometer
(spectral resolution of $45 \, \mu eV$) to selectively suppress
and filter the distinct components of the Mollow triplet. As
schematically depicted in Fig.1 (d) the first Michelson
interferometer is set to the distinct path difference for which
the two Mollow sidebands interfere constructively whereas the
central Rayleigh peak is suppressed via destructive interference :
$d = \lambda (\lambda + \Delta \lambda)/(2 \Delta \lambda)$. In a
second similar filtering step either one of the Mollow sidebands
is subsequently suppressed. A pre-condition of the applicability
of this method is that the path difference between the two
interferometer arms $l_{path}$ is well below the coherence length
$l_{coh}$ of the photons, which is readily fulfilled for our
experimental conditions where $l_{coh}^{min} \approx 71 \, mm \gg
l_{path}^{max} \approx 14 \, mm$. Figure 1 (a)-(c) illustrates the
functionality of the two-step double Michelson interferometer
filtering process via
high-resolution Mollow tiplet data.\\

% insert Fig. 2 here

Figure 2 shows a Mollow triplet power series of the QD under
investigation. Because of the Fabry-P\'{e}rot-based technique in
HRPL, all spectral features appear periodically with an offset
equal to the free spectral range of the interferometer ($\Delta
E_{FSR} = 62.04 \, \mu eV$). With increasing excitation power, the
evolution of two symmetrically spaced sidebands is clearly visible
(black arrows) around the central line. To extract the evolution
of coherence time $T_2$ of the Mollow sidebands a Lorentzian line
shape has been fitted to the spectra. The Inset of Fig.2 shows the
FWHM of the Mollow sidebands plotted versus the squared Rabi
frequency $\Omega^2$, from which the expected linear increase due
to excitation-induced dephasing is clearly observable
\cite{Ulrich.Ates:2011,Roy.Hughes:2011}. The herefrom calculated
$T_2$-times are depicted in the same plot. For the experimental
conditions of the two-photon interference measurements $\Omega =
45.5 \, \mu eV$ and $\Omega = 53.8 \, \mu eV$, $T_2$ is found to
be $264
\pm 35 \, ps$ and $237 \pm \, ps$, respectively. \\

% insert Fig. 3 here

Figure 3 (I-III) presents the core of our investigations, i.e.
Hong-Ou-Mandel type two-photon interference measurements on the
individual Mollow triplet sidebands together with the
auto-correlation measurements of the respective transition lines.

Figure 3 (I) shows the data set for a Rabi-splitting of $\Omega =
45.5 \, \mu eV$. The auto-correlation on the Mollow central peak
is expected to reveal Poissonian statistics, due to the fact, that
the emission of a Rayleigh photon does not modify the
dressed-state population
\cite{Schrama.Nienhuis:1992,Nienhuis:1993}. As is clearly visible
the Poissonian statistics is superimposed with a long-time
bunching, as commonly observed due to "blinking" of the excitonic
state between two or more neighboring competing states
\cite{Santori.Pelton:2001}. The phenomenon of "blinking" under
pure-resonant excitation can be assigned to the presence of a
competing QD exciton spin configuration, the non-radiative dark
excitonic state \cite{Bayer.Ortner:2002}. The auto-correlation
data has been fitted with a bidirectional exponential fit
revealing a decay timescale of $\tau_{bunch} = 9.51 \, ns$. Worth
to note this power-dependent bunching is observed to superimpose
all photon correlation measurements. High quality
indistinguishability measurements crucially rely on the purity of
single-photon emission, therefore auto-correlation measurements on
the red Mollow sideband have been performed in advance, revealing
$g^{(2)}(0)_{deconv.} = 0.12 \pm 0.03$ (considering the
time-resolution of the setup of $400 \, ps$). The correlation data
unambiguously demonstrates single-photon emission with a fitted
decay time constant of $\tau_r = 540 \, ps$. The finite background
in the auto-correlation can be attributed to a small drift of the
Michelson interferometer during the correlation measurement,
allowing the spectrally close Mollow central peak to contribute to
the detected signal as a non-vanishing uncorrelated background.

The Hong-Ou-Mandel type two-photon interference measurements shown
in Fig. 3 (I) (c) and (d) have been performed under resonant cw
excitation with an asymmetric fiber-based Mach-Zehnder
interferometer (fixed interferometer delay $\Delta \tau = 13 \,
ns$). By a manually adjustable polarizer inserted into one
interferometer arm the mutual photon polarization between the two
inputs of arms entering the Hong-Ou-Mandel beamsplitter can be set
orthogonal and therefore distinguishable where one expects
$g^{(2)}(0) = 0.5$. In parallel configuration the paths are
indistinguishable and $g^{(2)}(0)$ should ideally become zero. The
two-photon interference visibility calculated from the two
different configurations is defined as $V_{HOM}(\tau) =
[(g^{(2)}_{\perp}(\tau) -
g^{(2)}_{\parallel}(\tau))]/g^{(2)}_{\perp}(\tau)$
\cite{Patel.Bennett:2008,Ates.Ulrich:2009}.

The HOM data has then been modeled according to Ref.
\cite{Patel.Bennett:2010}. For this analysis the reflectivities R
and transmittivities T of the beamsplitters have been measured to
be $R = T = 0.5 \pm 0.02$ and the wavepacket overlap at the
Hong-Ou-Mandel beamsplitter as $V = 0.98 \pm 0.02$.  For fitting
the data we use $T_2 = 300 \, ps$ close to the extracted value of
$T_2 = 264 \pm 35 \,  ps$, as well as $\tau_{bunching}$ and
$\tau_r$ as derived from the auto-correlations discussed above.
$g^{(2)}(0)$ for the parallel HOM correlation measurement is
assumed to be slightly better than in (b) and (c), which was found
to be related to a more stable positioning of the Michelson
interferometers during the measurement time (less background from
the Mollow central line). Considering the setup response we find
$g^{(2)}_{\perp}(0)_{deconv.} = 0.61 \pm 0.04$ and
$g^{(2)}_{\parallel}(0)_{deconv.} = 0.04 \pm 0.01$, respectively.
Therefore, a close to ideal post-selected visibility of $V_{HOM} =
0.93 \pm 0.01$ (convoluted $V_{HOM}= 0.47 \pm 0.08$)
for the indistinguishability of the red Mollow sideband can be concluded.\\

Two-photon interference revealing indistinguishable photons has
also been verified for a larger Rabi splitting of $\Omega = 53.8
\, \mu eV$ on the red and blue Mollow sideband, respectively. The
measurements and evaluations are done similarly as described
above. Fig. 3 (II) depicts the data for red Mollow sideband
together with the auto-correlation on the central Mollow peak
revealing a long-timescale bunching with $\tau_{bunch} = 14.3 \,
ps$. The auto-correlation shows almost background-free
single-photon emission with a deconvoluted value of
$g^{(2)}(0)_{deconv.} = 0.08 \pm 0.02$. The timescale of the decay
is found to be $\tau_r = 570 \, ps$ and the extracted coherence
time derived from the HRPL power series of $T_2 = 237 \, ps$ fits
the data perfectly. Again a very high two-photon
visibility of $V_{HOM} = 0.97 \pm 0.02$ (convoluted $V_{HOM} = 0.43 \pm 0.07$) has been extracted from the deconvoluted fits to the Hong-Ou-Mandel data.\\

Figure 3 (III) shows the correlation data of the blue sideband.
$\tau_r$ and $T_2$ revealing full consistency with the
measurements on the red Mollow sideband at the same Rabi
splitting. The $g^{(2)}(0)_{deconv}$ is again found to be almost
background free with a value of $0.03 \pm 0.01$. The
Hong-Ou-Mandel visibility is
calculated to be $V_{HOM} = 0.92 \pm 0.02$ (convoluted $V_{HOM} = 0.39 \pm 0.08$).\\

It is worth to mention that the decay time $\tau_r$ extracted from
the auto-correlation measurements on the individual Mollow
sidebands is found to be smaller than the theoretically predicted
modified emission time of $\frac{\tau_{rad}}{(c^4 + s^4)} =
\tau_{mod}$ for a resonantly exited quantum dot in the
dressed-state regime. The two coefficients $c$ and $s$ are defined
as $c = \sqrt{\frac{\Omega^{'} + \Delta}{2 \Omega^{'}}}$ $s =
\sqrt{\frac{\Omega^{'} - \Delta}{2 \Omega^{'}}}$, where
$\Omega^{'} = \sqrt{\Omega^2 + \Delta ^2}$ is the effective Rabi
frequency and $\Delta$ the laser-detuning from resonance
\cite{Schrama.Nienhuis:1992}. For out experimental conditions of
strictly resonant excitation $\Delta = 0$ we find $s = c =
\frac{1}{\sqrt{2}}$. Independent time-resolved photoluminescence
investigations under quasi-resonant emitter-state preparation have
revealed a radiative lifetime of $\tau_{rad} = 1008 \, ps$,
therefore the modified decay time for the system is calculated to
be $\tau_{mod} = 2016 \, ps$. We emphasize that a similar
discrepancy between experimentally extracted $\tau_r$ to
theoretically calculated $\tau_{mod}$ has already been observed
previously \cite{Ulhaq.Weiler:2012} and might be attributed to the
distinctly different experimental conditions of cw resonant
excitation to pulsed quasi-resonant emitter-state preparation.
Nevertheless a profound explanation has to be left open for
further in-depth analysis.

In conclusion, we have presented a detailed study of the
indistinguishably of photons emitted from the Mollow triplet
sidebands from a single semiconductor QD. Independently measured
power series for emission coherence time extraction and
auto-correlation measurements on the Mollow triplet central and
sideband channels reveal full consistency with the cw
indistinguishability measurements that exhibit high two-photon
interference visibilities of up to $97 \%$.\\

We thank W.-M. Schulz for expert sample preparation, M. Heldmaier
and A. Ulhaq for fruitful discussions. We acknowledge financial
support of the DFG via the project MI500/23-1. S. Weiler
appreciates funding from the Carl-Zeiss Stiftung.

%%%%%%%%%%%%%%%%%%%%%%%%%%%%%%%%% References %%%%%%%%%%%%%%%%%%%%%%%%%%%%%%%%%%%%%%%%%%%

%\def\etal{\emph{et al.}}
%\bibliographystyle{apsrev}
%\bibliography{PRL_Ates_etal}

%%%%%%%%%%%%%%%%%%%%%%%% FIGURES (submission format) %%%%%%%%%%%%%%%%%%%%%%%%%%%%%%%%%%%

\newpage

\textbf{Figures}

\vspace{0.5cm}

\textbf{Figure~1, S. Weiler}

\begin{figure}[!h]
\label{Figure1}
\begin{center}
\includegraphics[width=8.0cm]{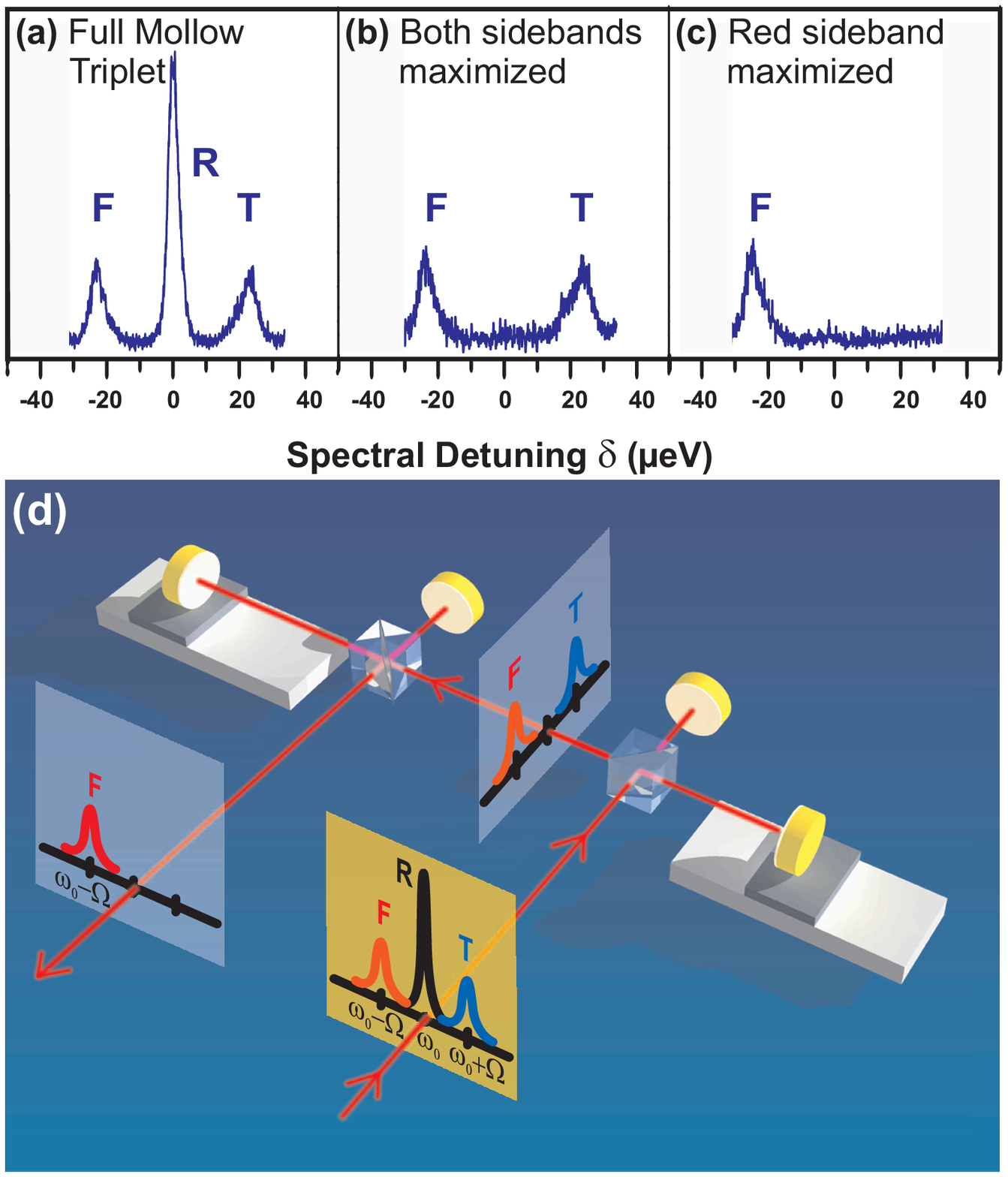}
\vspace{-0.3cm}
\end{center}
\end{figure}

\newpage
\textbf{Figure~2, S. Weiler}

\begin{figure}[!h]
\label{Figure2}
\begin{center}
\includegraphics[width=8.0cm]{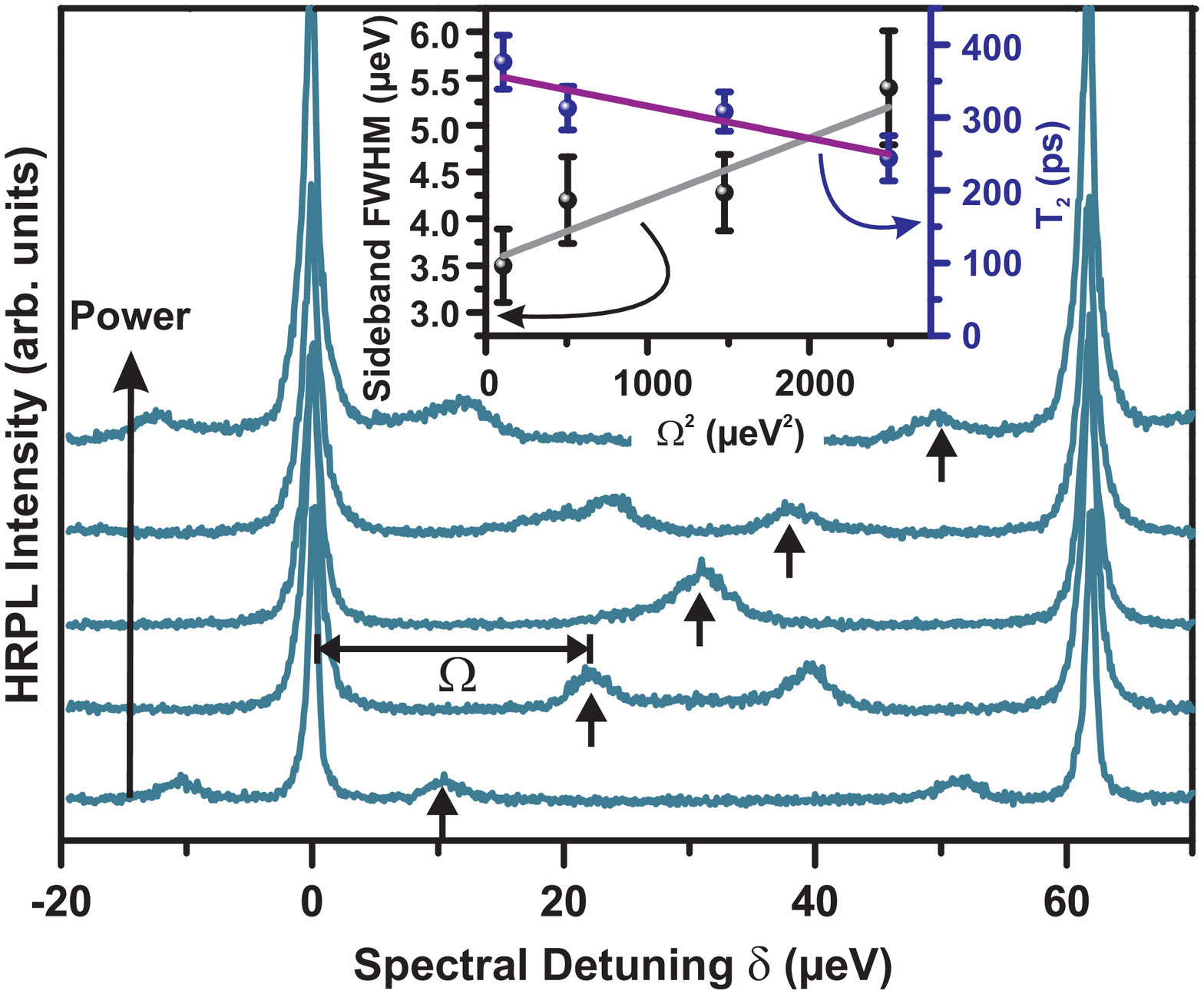}
\vspace{-0.3cm}
\end{center}
\end{figure}

\newpage
\textbf{Figure~3, S. Weiler}

\begin{figure}[!h]
\label{Figure3}
\begin{center}
\includegraphics[width=8.0cm]{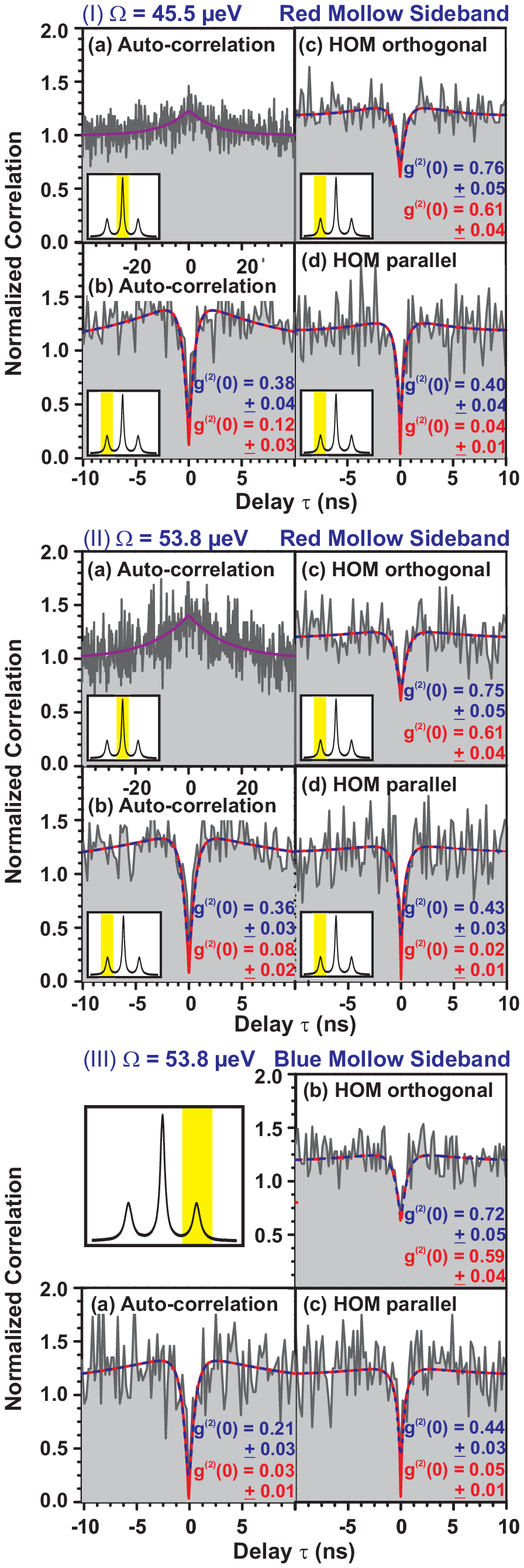}
\vspace{-0.3cm}
\end{center}
\end{figure}

%%%%%%%%%%%%%%%%%%%  FIGURE CAPTIONS (submission format) %%%%%%%%%%%%%%%%%%%%%%%%%%%%%%%

\newpage
\textbf{Figure Captions}

\textbf{FIG.1}

\textbf{Double Michelson Filtering Technique:} High-resolution
spectrum of (a) The full Mollow triplet without Michelson
filtering (b) The Mollow triplet sidebands (suppressed Rayleigh
line after first filtering step) (c) Only the red Mollow sideband
(after second filtering step). (d) Illustration of the double
Michelson interferometer setup used to spatially separate one
Mollow component from the other two.

\vspace{1cm}

 \textbf{FIG.2}

\textbf{Coherence Time Extraction:} Power dependent Mollow triplet
spectra. The belonging sidebands are marked with black arrows, the
additional peaks reflect the different orders of interference of
the scanning Fabry-P\'{e}rot interferometer. Inset: FWHM of the
Mollow sidebands, extracted from Lorentzian fits to the HRPL data
and the herefrom calculated coherence time $T_2$ versus
$\Omega^2$. \vspace{1cm}

 \textbf{FIG.3}

\textbf{Correlation Measurements:} Blue dashed (red straight) line
convoluted (deconvoluted) fits to the $g^{(2)}(\tau)$ data.

(I) Data set for the two-photon interference measurements of the
red Mollow sideband for $\Omega = 45.5 \mu eV$. (a)
Auto-correlation of the central Mollow peak revealing long-term
bunching with a timescale of $\tau_{bunch} = 9.51 \, ns$. (b)-(d)
Auto-correlation, perpendicular and parallel Hong-Ou-Mandel
measurements on the photon emission of the red Mollow sideband.
The fits reveal a modified emission time of $\tau_r = 540 \, ps$
and $T_2 = 300 \, ps$ close to the independently derived coherence
time of $264 \, ps$ from the HRPL spectra. The orthogonal HOM data
is in full consistency with these values, for the parallel case
$g^{(2)}(0)$ has been assumed to be slightly better than for the
other measurements. The deconvoluted values reveal a high degree
of two-photon visibility $V_{HOM} = 0.93 \pm 0.01$ (convoluted
$V_{HOM} = 0.47 \pm 0.08$).

(II) Similar data set than in (I) but at for $\Omega = 53.8 \, \mu
eV$. The values for fitting all correlation measurements are found
to be $\tau_{bunch} = 14.3 \, ps$, $T_2 = 237 \, ps$, $\tau_r =
570 \, ps$, exhibiting a visibility of $V_{HOM} = 0.97 \pm 0.02$
(convoluted $V_{HOM} = 0.43 \pm 0.07$).

(III) Auto- and HOM correlations for the blue Mollow sideband at
$\Omega = 53.8 \, \mu eV$. The fitting values $\tau_r =  570 \,
ps$ and $T_2 = 237 \, ps$ are in accordance with all correlation
measurements. Here the evaluations reveals $V_{HOM} = 0.92 \pm
0.02$ (convoluted $V_{HOM} = 0.39 \pm 0.08$).

\end{document}